\documentstyle[aps,prl,preprint,tighten]{revtex}

\newcommand{\half}{{\textstyle{1\over2}}}
\newcommand{\halfhalf}{(\half,\half)}
\newcommand{\halfzero}{(\half,0)}
\newcommand{\zerohalf}{(0,\half)}
\newcommand{\halfone}{(\half,1)}
\newcommand{\onehalf}{(1,\half)}
\newcommand{\feynslash}[1]{#1\hspace{-9pt}\slash\hspace{6pt}}

\newcommand{\tfrac}[2]{{\textstyle{\frac{#1}{#2}}}}

\renewcommand{\d}{{\rm d}}
\newcommand{\Tr}{\mbox{Tr\,}}
\newcommand{\Det}{\mbox{{\rm Det}}}
\newcommand{\beq}{\begin{equation}}
\newcommand{\eeq}[1]{\label{#1}\end{equation}}
\newcommand{\bea}{\begin{eqnarray}}
\newcommand{\eea}[1]{\label{#1}\end{eqnarray}}

\newcommand{\fft}[2]{{#1 \over #2}}
\newcommand{\ft}[2]{{\textstyle{{\scriptstyle #1}\over {\scriptstyle #2}}}}
\newcommand{\dalemb}[2]{{\vbox{\hrule height .#2pt
        \hbox{\vrule width.#2pt height#1pt \kern#1pt
                \vrule width.#2pt}
        \hrule height.#2pt}}}
\newcommand{\square}{\mathord{\dalemb{6.8}{7}\hbox{\hskip1pt}}}
\newbox\bstrutbox
\setbox\bstrutbox=\hbox{\vrule height12.5pt depth7.5pt width0pt}
\def\bstrut{\relax\ifmmode\copy\bstrutbox\else\unhcopy\bstrutbox\fi}

\begin{document}
\preprint{\vtop{\hbox{MCTP-02-61}\hbox{PUPT-2061}\hbox{hep-th/0211183}\vskip24pt}}

\title{Quantum discontinuity for massive spin $3/2$ with a $\Lambda$ term}

\author{M.~J.~Duff$\,{}^{1,*}$, James T.~Liu$\,{}^{1,2,*}$ and
H.~Sati$\,{}^{1,}$%
\footnote{Email addresses: \{mduff, jimliu, hsati\}@umich.edu}}

\address{${}^1$Michigan Center for Theoretical Physics\\
Randall Laboratory, Department of Physics, University of Michigan\\
Ann Arbor, MI 48109--1120, USA}

\address{\strut\\
${}^2$Department of Physics, Princeton University, Princeton, NJ 08544,
USA}

\maketitle

\begin{abstract}
We show that the recently demonstrated absence of the van
Dam-Veltman-Zakharov discontinuity for massive spin $3/2$ with a $\Lambda$
term is an artifact of the tree approximation, and that the discontinuity
reappears at one loop. As a numerical check on the
calculation, we rederive the vanishing of the one-loop beta function
for $D=11$ supergravity on $AdS_{4} \times S^{7}$ level-by-level in the
Kaluza-Klein tower.

\end{abstract}


\ifpreprintsty\else
\begin{multicols}{2}
\fi
\narrowtext


\section{Introduction}

An old question is whether the graviton has exactly zero mass or
perhaps a small but non-zero mass.  This issue seemed to have been
resolved by van Dam and Veltman \cite{vdv} and, independently,
Zakharov \cite{zak} when they noted that there is a discrete difference
between the propagator of a strictly massless graviton and that of a
graviton with mass $M$ in the $M\rightarrow 0$ limit.  The massless
limit of a massive graviton then yields a bending of light by the sun
which is only 3/4 of the Einstein value.  A similar discontinuity
appears in the massless limit of a massive spin $3/2$ \cite{Stelle}.

Subsequently, however, these discontinuities were called into question in
\cite{Kogan,Porrati} for the case of the graviton and in
\cite{grassi,Deser1} for the case of the gravitino, by pointing out 
that they disappear if the background spacetime is anti-de Sitter 
(AdS) or more generally Einstein, satisfying
\begin{equation}
\label{Einstein}
R_{\mu\nu} = \Lambda g_{\mu\nu},
\end{equation}
with a non-zero cosmological constant $\Lambda\neq0$ provided
$M^2/\Lambda\to0$.

Yet in recent work \cite{dilkes} we have shown that the above disappearance
of the discontinuity for the massive graviton is an artifact of the
tree level approximation and that the discontinuity reappears at one
loop as a result of going from five degrees of freedom to two%
\footnote{Similar classical continuity but quantum discontinuity arises
in the partially massless \cite{Waldron} limit as a result of going from
five degrees of freedom to four \cite{partially}.}.
In this paper we exhibit a similar one-loop discontinuity for the massive
gravitino, as a result of going from four degrees of freedom to two.

That a cosmological constant cannot cure the spin $3/2$ discontinuity
at the one-loop level is an observation that could have been made in
1978.  Then it was shown that the gravitational axial anomaly for a
Rarita-Schwinger particle is $-21$ times that of a Dirac particle in the
massless case, but $-20$ times in the massless limit of the massive case
\cite{CD1,CD2}.  Since the axial anomaly depends only on the Pontryagin 
density $R\,{}^{*}\!R$, moreover, it is completely unaffected by the 
presence of a cosmological constant.

The quantization of a single massive spin $3/2$ field in the presence of
a cosmological constant is treated in section \ref{massive} using the
St\"uckelberg formalism~\cite{stuck,grassi} which introduces a massive
spin $3/2$ gauge invariance through the addition of an auxiliary spin $1/2$
field.  This approach allows us to carry over many of the same 
procedures used for a massless spin $3/2$ field, and furthermore allows 
a simple comparison between massive and massless cases.  In section 
\ref{loop} we compute the one-loop partition function and verify that 
a quantum spin $3/2$ discontinuity arises.

In a previous paper on spin $2$ \cite{dilkes}, we made the important caveat
that our results about quantum discontinuities apply to theories where
the gauge symmetry is broken explicitly by the addition of a
Pauli-Fierz mass term.  We were agnostic about whether the problem is
avoided if the graviton gets its mass through a dynamical mechanism of
the kind appearing in the Karch-Randall braneworld
\cite{Karch,Porrati:2001db}, since this requires a separate
treatment.  A similar caveat applies to the spin $3/2$ case.
The supersymmetric Karch-Randall mechanism, which includes the case of spin
$3/2$, is discussed in \cite{Duff:2002ab} from the braneworld point of 
view and we mention it again briefly in section \ref{conclusion}, 
postponing a more thorough analysis of the VVZ problem to a future
publication.

In the spin $3/2$ case there is a third possibility: the gravitino may
acquire a mass through a conventional super-Higgs effect 
\cite{Deser2,Volkov,Cremmer1,Cremmer2}.  Here, as described in section 
\ref{loop}, the massive theory does limit smoothly to a massless 
theory with the same number degrees of freedom.  In particular, the 
axial anomaly suffers no discontinuity.  All this is true whether or 
not there is a cosmological constant. However, even for simple 
supergravity the massless theory is not the minimal one so the spin 
$3/2$ analogue of the ``wrong bending of light'' feature continues to 
apply.  Similarly, the dynamical mechanism for a massive spin $2$ 
suggested in \cite{dynamic} has a smooth massless limit, but it limits 
to a tensor-scalar theory with the wrong bending of light.  The 
Karch-Randall mechanism, on the other hand, is claimed to be both 
continuous and to limit to Einstein gravity \cite{Katz}.

Finally, in section
\ref{super}, we collect the results for fields of spins $\le2$ and
examine the implication for massive supermultiplets.  A check on the
numerical calculations, which is of interest in its own right, is
provided by the massive Kaluza-Klein tower arising from the
$AdS_{4}\times S^{7}$ compactification of $D=11$ supergravity.  We
confirm that the one-loop beta function vanishes not only for the
massless modes \cite{Gibbons} but also for the massive Kaluza-Klein
tower level by level \cite{GN,Inami}.


\section{St\"uckelberg quantization of massive spin 3/2}
\label{massive}

We work in four dimensions with Euclidean signature \mbox{$(+,+,+,+)$}
and a cosmological constant $\Lambda$.  As in \cite{grassi,Deser1}, 
our starting point is the Rarita-Schwinger action for a massive 
spin $3/2$ field coupled to a source
\begin{equation}
e^{-1}{\cal L}_{3/2} =
-\frac{1}{2}\bar{\Psi}_{\mu}\gamma^{\mu\rho\nu}\nabla_{\rho}\Psi_{\nu}
+\fft{M}2\bar{\Psi}_{\mu}\gamma^{\mu\nu}\Psi_{\nu}+\bar{\Psi}^\mu J_\mu.
\label{eq:32act}
\end{equation}
In flat (or Ricci-flat) space, the massless equation (obtained by setting
$M=0$) is invariant under the spin $3/2$ gauge transformation
\begin{equation}
\delta \Psi_{\mu}= \nabla_{\mu} \epsilon,
\end{equation}
provided the source is conserved, $\nabla\cdot J=0$.  Gauge fixing and
quantization of this system was previously carried out in \cite{das}.

In the presence of a cosmological constant, ``masslessness'' (in the
sense of propagating reduced degrees of freedom) is no longer given by $M=0$,
but rather by $M^2=-\Lambda/3$.  Note that this is only possible in AdS;
for future convenience we take $\Lambda=-3\overline{m}^2$ when
specializing to the AdS case (so that masslessness corresponds to
$M^2=\overline m^2$).  For the massless case, the combined action is
invariant under the gauge transformation
\begin{equation}
\delta \Psi_{\mu}= {\cal D}_\mu\epsilon
\equiv (\nabla_{\mu} + \half M  \gamma_{\mu}) \epsilon,
\label{gauge}
\end{equation}
which, in the language of supergravity, is simply a supersymmetry
transformation on the gravitino.  However other values of mass,
$M^2\ne-\Lambda/3$, break this gauge invariance.  To see this,
we note that the Lagrangian (\ref{eq:32act}) may be written in terms
of the ``supercovariant derivative'' ${\cal D}_\mu$ as simply
\begin{equation}
e^{-1}{\cal L}_{3/2}=-\fft12\bar{\Psi}_\mu\gamma^{\mu\rho\nu}{\cal
D}_\rho\Psi_\nu+\bar{\Psi}^\mu J_\mu.
\end{equation}
The variation under the transformation (\ref{gauge}) is
\begin{eqnarray}
\delta(e^{-1}{\cal L}_{3/2})&=&
-\fft14\bar{\Psi}_\mu\gamma^{\mu\rho\nu}
[{\cal D}_\rho,{\cal D}_\nu]\epsilon + \fft14\overline{
[{\cal D}_\rho,{\cal D}_\mu]\epsilon}\gamma^{\mu\rho\nu}\Psi_\nu
+\overline{{\cal D}^\mu\epsilon}J_\mu\nonumber\\
&=&\fft14(\bar\Psi^\mu\gamma^\nu\epsilon
-\bar\epsilon \gamma^\mu\Psi^\nu)(3M^2g_{\mu\nu}-R_{\mu\nu}+\ft12g_{\mu\nu}R)
-\bar\epsilon{\cal D}^\mu J_\mu.
\end{eqnarray}
Substituting in the Einstein condition, $R_{\mu\nu}=\Lambda g_{\mu\nu}$,
but keeping $M$ and $\Lambda$ independent, this turns out to be
\begin{equation}
\delta(e^{-1}{\cal L}_{3/2}) =\fft{\Lambda+3M^2}4
(\bar\Psi\cdot\gamma\epsilon -\bar\epsilon \gamma\cdot\Psi)
-\bar\epsilon{\cal D}^\mu J_\mu.
\end{equation}
This demonstrates that gauge invariance of the action demands both
masslessness, $M^2=-\Lambda/3$, and supercovariant conservation of the
current, ${\cal D}^\mu J_\mu=0$.


Although gauge invariance is lost for $M^2\ne-\Lambda/3$, it may be
restored using a St\"uckelberg approach \cite{stuck}.  In the present
case, this amounts to the introduction of an auxiliary spin $1/2$ field,
$\chi$, transforming as
\begin{equation}
\delta\chi=\epsilon,
\end{equation}
so that the shifted quantity $\Psi_{\mu}'\equiv\Psi_{\mu}-{\cal D}_\mu \chi$
remains gauge invariant.  Making the replacement
$\Psi_\mu\to\Psi_\mu'$ in (\ref{eq:32act}), we find
\begin{eqnarray}
e^{-1}{\cal L}_{3/2}&=&
-\fft12\bar{\Psi}_{\mu} \gamma^{\mu \rho \nu} \nabla_{\rho} \Psi_{\nu}
+\fft{M}2 \bar{\Psi}_{\mu}\gamma^{\mu \nu}\Psi_{\nu}
+\bar\Psi^\mu J_\mu\nonumber\\
&&-{{\Lambda + 3M^2}\over {4}} [ \bar{\chi} (\feynslash{\nabla} + 2M)\chi +
\bar{\Psi} \cdot \gamma \chi - \bar\chi \gamma \cdot \Psi]
+\bar\chi{\cal D}_\mu J^\mu.
\label{eq:slag}
\end{eqnarray}
While we have avoided explicit use of supergravity techniques, it is
welcome to see that the St\"uckelberg spinor $\chi$ may be
interpreted in a supergravity language as a Goldstino field associated
with broken supersymmetry; it is the field eaten by the gravitino to
provide its mass.  Indeed, scaling
$\chi\to\sqrt{2/(\Lambda+3M^2)}\chi$ to give a canonical kinetic term,
it is clear that $\chi$ couples to the gravitino via the Goldstino
current, $\bar\Psi^\mu J^{(\chi)}_\mu$ where
$J^{(\chi)}_{\mu}=\sqrt{(\Lambda+3M^2)/2}\gamma^{\mu} \chi$.  This
current clearly vanishes in the limit $M^2\to-\Lambda/3$.

For negative cosmological constant, this spin $3/2$ Higgs mechanism may be
interpreted in terms of a decomposition of AdS representations.  Recall
that for AdS$_4$, with isometry group $SO(3,2)$, representations are
labeled by $D(E_0,s)$, where $E_0$ is the minimum energy and $s$ is the
spin.  Generic unitary representations require $E_0\ge s+1$, with saturation
corresponding to massless (shortened) representations.  In particular, a
massless graviton transforms as a $D(3,2)$, while a massless gravitino
transforms as a $D(5/2,3/2)$.  The Higgs mechanism in AdS is then
expressed in terms of a massive representation becoming reducible in the
zero-mass limit
\begin{equation}
D(s+1+\epsilon,s) \rightarrow D(s+1,s) \oplus
D(s+2,s-1)\quad\hbox{as}\quad \epsilon \rightarrow 0,
\label{eq:adshig}
\end{equation}
or $D(5/2+\epsilon,3/2)\to D(5/2,3/2)+D(7/2,1/2)$ for the gravitino.
To examine the connection between this decomposition and the
St\"uckelberg Lagrangian (\ref{eq:slag}), we make use of the relations
between mass and $E_0$ for spin $1/2$ and spin $3/2$; these are
generalizations of the Breitenlohner-Freedman bound \cite{BF} for higher
spins and are given by $E_0=\frac{3}{2}+|m_{1/2}/\overline{m}|$
for spin $1/2$ and $E_0=\frac{3}{2}+ |m_{3/2}/\overline{m}|$ for
spin $3/2$ \cite{englert,Casher:1984ym}.
Reading off $m_{1/2} = 2 M$ and $m_{3/2} = M$
from (\ref{eq:slag}), and taking the massless limit
$M\to\overline{m}$, we indeed find the values $E_0=7/2$ and $E_0=5/2$
for spins $1/2$ and $3/2$, respectively.  This confirms that the mass terms
in the Lagrangian are appropriate to an AdS Higgs mechanism.

We now turn to the issue of gauge fixing and quantization of the
St\"uckelberg Lagrangian, (\ref{eq:slag}).  Before proceeding, we find
it convenient to make the field redefinition \cite{Endo:1985km}
\begin{equation}
\Psi_{\mu}=\phi_{\mu} - \frac{1}{2} \gamma_{\mu}\gamma \cdot \phi.
\label{eq:gtri}
\end{equation}
This is analogous to the decomposition of the linearized fluctuation
$h_{\mu\nu}$ of the graviton into ${\tilde h}_{\mu\nu}=h_{\mu\nu}
-\fft12g_{\mu\nu}h$.  Making this redefinition and performing
the aforementioned rescaling of $\chi$, we obtain
\begin{eqnarray}
e^{-1}{\cal L}_{3/2}&=&-\ft12\bar\phi^{\mu}(\feynslash{\nabla}+M)\phi_{\mu}
-\ft14\bar\phi \cdot \gamma (\feynslash{\nabla} -2M)\gamma \cdot \phi
-\ft12 \bar\chi(\feynslash{\nabla} + 2M)\chi\nonumber\\
&&+\fft12\sqrt{\fft{\Lambda + 3M^2}2}(\bar \phi
\cdot \gamma \chi - \bar \chi \gamma \cdot \phi)+\bar\phi^\mu
J_\mu-\ft12\bar\phi\cdot\gamma\gamma\cdot J+\sqrt{\fft2{\Lambda+3M^2}}
\bar\chi{\cal D}_\mu J^\mu.
\label{eq:newl}
\end{eqnarray}
Note that, because of the denominator in the last term, the supercurrent,
$J_\mu$, must be supercovariantly conserved in the massless limit.

It is instructive to examine the massless spin $3/2$ field in flat space.
In this case, (\ref{eq:newl}) becomes simply
\begin{equation}
e^{-1}{\cal L}_{3/2}=-\ft12\bar\phi^{\mu}\feynslash{\nabla}\phi_{\mu}
-\ft14\bar\phi \cdot \gamma \feynslash{\nabla}\gamma \cdot \phi
+\bar\phi^\mu J_\mu-\ft12\bar\phi\cdot\gamma\gamma\cdot J.
\end{equation}
There has been a long history in quantizing the massless spin $3/2$ field,
especially as it pertains to supergravity theories.  A convenient choice
of gauge fixing would be to simply introduce the term
\begin{equation}
e^{-1}{\cal L}_{\rm gf}=\fft14\bar\phi\cdot\gamma
\feynslash{\nabla}\gamma\cdot\phi,
\label{eq:zmgf}
\end{equation}
so that the resulting Feynman gauge Lagrangian has a Dirac-like form
\begin{equation}
e^{-1}\widetilde{\cal L}_{3/2}=-\fft12\bar\phi^\mu\feynslash\nabla\phi_\mu
+\bar\phi^\mu J_\mu-\ft12\bar\phi\cdot\gamma\gamma\cdot J.
\label{eq:zmlag}
\end{equation}
This gauge fixing of the spin $3/2$ field is complicated by the fact that,
in addition to the ordinary Faddeev-Popov ghosts, one must include a
Nielsen-Kallosh ghost related to $\det\feynslash{\nabla}$ which shows
up in the Gaussian gauge fixing procedure \cite{Nielsen:1978mp,Kallosh:de}.
This is avoided in the $\gamma\cdot\Psi=0$ gauge \cite{CD1,CD2}.

Returning to the massive spin $3/2$ field in a cosmological background,
we find that the appropriate generalization of the gauge fixing term
(\ref{eq:zmgf}) is
\begin{equation}
e^{-1}{\cal L}_{\rm gf}=-\fft{2M+\sqrt{M^2-\Lambda}}{8\sqrt{M^2-\Lambda}}
\overline{\gamma\cdot\phi+\alpha\chi}(\feynslash{\nabla}-\sqrt{M^2-\Lambda})
(\gamma\cdot\phi+\alpha\chi),
\label{eq:ggf}
\end{equation}
where
\begin{equation}
\alpha=(2M-\sqrt{M^2-\Lambda})\sqrt{2\over \Lambda+3M^2}.
\end{equation}
The coefficients here are chosen so that the gauge fixed Lagrangian
takes the simple quadratic form
\begin{eqnarray}
e^{-1}\widetilde{\cal L}_{3/2}&=&-\ft12\bar\phi_\mu(\feynslash{\nabla}+M)
\phi^\mu
-\ft12\bar\xi(\feynslash{\nabla}+\sqrt{M^2-\Lambda})\xi+\bar\phi^\mu
J_\mu -\ft12\bar\phi\cdot\gamma\gamma\cdot J\nonumber\\
&&+\fft\alpha2\sqrt{2\over\Lambda+3M^2}\bar\phi\cdot\gamma{\cal D}^\mu
J_\mu +\sqrt{2-\alpha^2\over\Lambda+3M^2}\bar\xi{\cal D}^\mu J_\mu,
\label{eq:qlag}
\end{eqnarray}
which may be compared to the massless flat space case of
(\ref{eq:zmlag}).  Here $\xi$ is a rescaled and shifted St\"uckelberg field
\begin{equation}
\xi=\sqrt{2\over2-\alpha^2}(\chi+\ft12\alpha\gamma\cdot\phi).
\end{equation}

We are now in a position to compute the gauge fixed propagator evaluated
between conserved sources.  The tree-level amplitude takes the form
\begin{eqnarray}
{\cal A}&=&\bar J_\mu\left(g^{\mu\alpha}-\fft12\gamma^\mu\gamma^\alpha
-\fft\alpha2\sqrt{2\over\Lambda+3M^2}\stackrel\leftarrow{{\cal D}^\mu}
\gamma^\alpha\right)
[\feynslash\nabla+M]^{-1}_{\alpha\beta}\nonumber\\
&&\qquad\qquad\times\left(g^{\beta\nu}-\fft12\gamma^\beta\gamma^\nu+\fft\alpha2
\sqrt{2\over\Lambda+3M^2}\gamma^\beta{\cal D}^\nu\right)J_\nu\nonumber\\
&&+\fft{2-\alpha^2}{\Lambda+3M^2}\bar J_\mu\stackrel\leftarrow{{\cal
D}^\mu}[\feynslash\nabla+\sqrt{M^2-\Lambda}]^{-1}{\cal D}^\nu J_\nu.
\label{eq:amp}
\end{eqnarray}
If we take current conservation to be ${\cal D}^\mu J_\mu=0$ for
arbitrary mass $M$, we find the simple result
\begin{equation}
{\cal A}=\bar J_\mu\fft1{\feynslash\nabla+M}J^\mu+\fft12\bar
J\cdot\gamma
\fft{\feynslash\nabla+2M}{\square-M^2}\gamma\cdot J,
\end{equation}
which has a smooth limit, both for $M^2\to 0$ and for
$M^2\to-\Lambda/3$ \cite{Deser1}.  However, as explained in \cite{Deser1},
we are certainly not allowed to modify the current conservation
equation, provided we wish to make a proper comparison with the massless
theory.

For comparison to the massless AdS limit, we note that proper current
conservation takes the form $\overline{\cal D}^\mu J_\mu\equiv
(\nabla^\mu+\fft{\overline m}2\gamma^\mu)J_\mu=0$.
In this case, we find ${\cal D}^\mu
J_\mu=\fft12(M-\overline{m})\gamma\cdot J$.  Substituting this into
(\ref{eq:amp}), and noting that the squares of the Dirac operators are
\begin{eqnarray}
-\Delta_{(1/2)}&\equiv&[\feynslash\nabla\!\!_{(1/2)}]^2
=\square-\fft14R,\nonumber\\
-\Delta_{(3/2)}&\equiv&[\feynslash\nabla\!\!_{(3/2)}]_{\mu\nu}^2
=(\square-\fft14R)g_{\mu\nu}
+\fft12R_{\mu\nu\alpha\beta}\gamma^{\alpha\beta},
\label{eq:dops}
\end{eqnarray}
we now find
\begin{equation}
{\cal A}=\bar J_\mu\fft1{\feynslash\nabla+M}J^\mu+\bar J\cdot\gamma
\fft{(M+2\overline{m})\feynslash\nabla+(M^2+2M\overline{m}+3\overline{m}^2)}
{3(M+\overline{m})(\square-M^2)}\gamma\cdot J,
\end{equation}
which agrees with \cite{Deser1}, and contains the spin $3/2$ version of
the van Dam-Veltman-Zakharov discontinuity \cite{grassi,Deser1}.  
Namely, in the case where both $M$ and $\overline{m}$ approach zero, 
we have
\begin{equation}
{\cal A}\sim\bar J_\mu\fft1{\feynslash\nabla}J^\mu+\fft{M+2\overline{m}}
{3(M+\overline{m})}\bar J\cdot\gamma\fft1{\feynslash\nabla}\gamma\cdot J,
\end{equation}
which demonstrates the sensitivity to the order of limits.  In particular,
taking the massless AdS limit, $M\to\overline{m}$, we recover the expected
factor of $1/2$ in the second term relative to the first.  But taking
the flat space limit, $\overline{m}\to0$, we find instead the
discontinuous factor of $1/3$.
%


\section{The one-loop partition function}
\label{loop}

The gauge fixing term (\ref{eq:ggf}), corresponding to the condition
$\gamma\cdot\phi+\alpha\chi=b$ (with $b$ a constant spinor), must be
accompanied by a pair of Faddeev-Popov ghosts with action connected to
the variation
\begin{equation}
\delta_\epsilon(\gamma\cdot\phi+\alpha\chi)=-\gamma^\mu\delta_\epsilon\Psi_\mu
+\alpha\delta_\epsilon\chi=-(\feynslash\nabla+\sqrt{M^2-\Lambda})\epsilon.
\end{equation}
In addition, there is a Nielsen-Kallosh ghost
\cite{Nielsen:1978mp,Kallosh:de} yielding the determinant of
$\feynslash\nabla-\sqrt{M^2-\Lambda}$, as can be read off directly from
(\ref{eq:ggf}).  Given these considerations in the ghost sector, it is
now possible to compute the one-loop partition function for the massive
spin $3/2$ field.  Collecting together the spin $3/2$ field $\phi_\mu$ and the
St\"uckelberg spinor $\xi$ with Lagrangian given by (\ref{eq:qlag}),
as well as the Faddeev-Popov and Nielsen-Kallosh ghosts, the one-loop
partition function becomes
\begin{equation}
Z[\Psi]  \propto
{{\Det_{\rm red}[\feynslash\nabla\!\!_{(3/2)}+M]
\Det[\feynslash\nabla\!\!_{(1/2)} +\sqrt{M^2-\Lambda}]}
\over
{\Det^2[\feynslash\nabla\!\!_{(1/2)}+\sqrt{M^2-\Lambda}]
\Det[\feynslash\nabla\!\!_{(1/2)}-\sqrt{M^2-\Lambda}]}},
\label{eq:olpart}
\end{equation}
where $\feynslash\nabla\!\!_{(3/2)}$ acts on {\it reducible}
states $\phi_\mu$.

Evaluation of the fermion determinants is most conveniently performed by
first squaring the Dirac operators corresponding to spin $1/2$   and 
spin-3/2 fields.  For Dirac fermions, we have simply
\begin{eqnarray}
\Det[\feynslash\nabla\!\!_{(1/2)} + \sqrt{M^2-\Lambda}]
&=&\Det^{\fft12}[\feynslash\nabla\!\!_{(1/2)} +\sqrt{M^2-\Lambda}]
\Det^{\fft12}[\feynslash\nabla\!\!_{(1/2)}-\sqrt{M^2-\Lambda}] \nonumber\\
&=&\Det^{\fft12}[{\Delta\halfzero}+\Delta\zerohalf + M^2-\Lambda],
\end{eqnarray}
and
\begin{eqnarray}
\Det_{\rm red}[\feynslash\nabla\!\!_{(3/2)} + M]
&=&\Det_{\rm red}^{\fft12}[\feynslash\nabla\!\!_{(3/2)}+ M]
\Det_{\rm red}^{\fft12}[\feynslash\nabla\!\!_{(3/2)}-M] \nonumber\\
&=&\Det_{\rm red}^{\fft12}[{\Delta\halfone}+\Delta\onehalf + M^2],
\end{eqnarray}
where $\Delta(A,B)$ are the Laplacians acting on irreducible $(A,B)$ 
representations of the Lorentz group \cite{CD1,CD2}.  In particular, 
$\Delta\halfzero$ and $\Delta\halfone$ are the chiral components of 
the operators given in (\ref{eq:dops}).  In terms of these 
determinants, the partition function, (\ref{eq:olpart}), becomes
\begin{eqnarray}
Z[\Psi]&\propto&
\Det_{\rm red}^\fft12 \Bigl[\Delta_{(3/2)} + M^2 \Bigl]
\Det^\fft12 \Bigl[\Delta_{(1/2)} + M^2-\Lambda \Bigl]\nonumber\\
&&\times\Det^{-1}\Bigl[\Delta_{(1/2)}+ M^2-\Lambda\Bigl]
\Det^{-\fft12}\Bigl[\Delta_{(1/2)} + M^2-\Lambda \Bigl].
\label{amplitude}
\end{eqnarray}
Note that the appropriate operators are given by the non-chiral combinations
$\Delta_{(1/2)} = \Delta\halfzero+{\Delta\zerohalf}$ and
$\Delta_{(3/2)} = \Delta\halfone+{\Delta\onehalf}$.  Because the ghost
and St\"uckelberg determinants have the same form, there is a partial
cancellation in the partition function.  In addition, we note from
(\ref{eq:dops}) that $\Delta_{(3/2)}=\Delta_{(1/2)}g_{\mu\nu}
-\fft12R_{\mu\nu\alpha\beta}\gamma^{\alpha\beta}$.  Thus when acting
on a pure gamma-trace field $\phi_\mu=\gamma_\mu\chi$, we find (in an
Einstein background)
\begin{equation}
\bar\chi\gamma_\mu[\Delta_{(3/2)}^{\mu\nu}+M^2g^{\mu\nu}]\gamma_\nu\chi=
4\bar\chi[\Delta_{(1/2)}+M^2-\Lambda]\chi,
\end{equation}
so that, up to a constant
\begin{equation}
\Det_{\rm red}\Bigl[\Delta_{(3/2)}+M^2\Bigr]=\Det\Bigl[\Delta_{(3/2)}+M^2\Bigr]
\Det\Bigl[\Delta_{(1/2)}+M^2-\Lambda\Bigr].
\end{equation}
where $\Delta_{(3/2)}$ on the right hand side acts on gamma-traceless
spin-3/2 fields, $\gamma\cdot\phi=0$.
As a result, the one-loop effective action for the spin-3/2 field
in an Einstein background takes on the simple form
\begin{equation}
\Gamma^{(1)}[\Psi] = - \ln Z[\Psi]
= -\tfrac{1}{2} \ln \Det \Bigl[\Delta_{(3/2)} + M^2 \Bigl]
+ \tfrac{1}{2} \ln \Det \Bigl[\Delta_{(1/2)} + M^2 -\Lambda\Bigl].
\end{equation}

For comparison, in the strictly massless case, $M^2=\overline{m}^2$,
there is no St\"uckelberg field, and the partition function has the form
\begin{equation}
Z[\Psi]_{\rm massless}  \propto
{{\Det_{\rm red}[\feynslash\nabla\!\!_{(3/2)} +\overline{m}]}
\over
{\Det^2[\feynslash\nabla\!\!_{(1/2)} + 2\overline{m}]
\Det[\feynslash\nabla\!\!_{(1/2)} - 2\overline{m}]}}.
\end{equation}
This leads instead to an effective action
\begin{equation}
\Gamma^{(1)}[\Psi]_{\rm massless}
= -\tfrac{1}{2} \ln \Det \Bigl[\Delta_{(3/2)} +\overline{m}^2\Bigl]
+ \ln \Det \Bigl[\Delta_{(1/2)} + 4\overline{m}^2\Bigl].
\end{equation}
The difference in these two expressions reflects the fact that in the
massive case the St\"uckelberg (or Goldstino) spinor provides an additional
two degrees of freedom to the gravitino.  This provides the correct
counting of degrees of freedom, namely two for the massless case and
four for the massive one.

The actual forms for the determinants of the above operators may be
computed in a heat-kernel expansion.  We focus on the coefficient functions
$b_k^{(\Delta)}$ in the expansion
\begin{equation}
\Tr e^{-\Delta t}
= \sum_{k=0}^\infty t^{(k-4)/2}
\int \d^4x \sqrt{g} \, b_k^{(\Lambda)},
\end{equation}
which were calculated in \cite{CD1,CD2} for the general operators 
$\Delta(A,B)$ and generalized in \cite{CD3,Gibbons} to allow for a 
cosmological constant.  In particular,
\begin{eqnarray}
180(4\pi)^2b_4\halfzero&=&-\frac{7}{4}R_{\mu\nu\rho\sigma}R^{\mu\nu\rho\sigma}
+12\Lambda^2 -\frac{15}{4}R_{\mu\nu\rho\sigma}{}^{*}
R^{\mu\nu\rho\sigma},\nonumber\\
180(4\pi)^2 b_4{\onehalf}&=&\fft{219}{4} 
R_{\mu\nu\rho\sigma}R^{\mu\nu\rho\sigma}-24\Lambda^2+
\frac{285}{4}R_{\mu\nu\rho\sigma}{}^{*}R^{\mu\nu\rho\sigma},
\end{eqnarray}
It is also easy to generalize the results (by 
summing over chiralities) to encompass the reducible Dirac 
combinations $\Delta_{(1/2)}$ and $\Delta_{(3/2)}$ that are of present 
interest.  For the $b_4$ coefficients, we find %
\begin{eqnarray}
180(4\pi)^2b_4^{(1/2)}&=&-\frac{7}{2}R_{\mu\nu\rho\sigma}R^{\mu\nu\rho\sigma}
+24\Lambda^2,\nonumber\\
180(4\pi)^2 b_4^{(3/2)}&=&\fft{219}{2}
R_{\mu\nu\rho\sigma}R^{\mu\nu\rho\sigma}-48\Lambda^2.
\end{eqnarray}

Given the above $b_4$ coefficients, it is simple to extend these results
to cover the relevant massive operators, $\Delta_{(3/2)}+M^2$ and
$\Delta_{(1/2)}+M^2-\Lambda$.  As was done in \cite{CD2,partially}, we
note that for constant $X$, the $b_4$ coefficient for the operator
$\Delta-X$ is given by the sum $b_4^{(\Delta-X)} =
b_4+Xb_2+\fft12X^2b_0$.  We obtain the resulting coefficients
\begin{eqnarray}
180(4\pi)^2 b_4^{(\Delta_{(1/2)}+M^2-\Lambda)}&=&
-\frac{7}{2} R_{\mu\nu\rho\sigma} R^{\mu\nu\rho\sigma} +144 \Lambda^2
- 480\Lambda M^2 + 360 M^4,\nonumber\\
180(4\pi)^2 b_4^{(\Delta_{(3/2)}+M^2-\Lambda)}&=&
\frac{219}{2} R_{\mu\nu\rho\sigma} R^{\mu\nu\rho\sigma} -48 \Lambda^2
+1440\Lambda M^2 +1080M^4.
\end{eqnarray}
These $b_4$ coefficients are perfectly smooth functions of $M^2$.  Thus
we may compare the results for the massive gravitino in the limit
$M^2\to-\Lambda/3$
\begin{eqnarray}
180(4\pi)^2b_4^{(\rm massive)}
&&= 180(4\pi)^2 \Bigl[b_4^{(\Delta_{(3/2)}+M^2)}
- b_4^{(\Delta_{(1/2)}+M^2-\Lambda)}\Bigr]\nonumber\\
&& \rightarrow
113 R_{\mu\nu\rho\sigma} R^{\mu\nu\rho\sigma} -752 \Lambda^2 \, ,
\end{eqnarray}
which clearly differs from the pure massless ($M^2=-\Lambda/3$) result
\begin{eqnarray}
180(4\pi)^2b_4^{(\rm massless)}&=& 180(4\pi)^2 \Bigl[
b_4^{(\Delta_{(3/2)}+\overline{m}^2)}
-2 b_4^{(\Delta_{(1/2)}+4\overline{m}^2)} \Bigr]\nonumber\\
&&= \frac{233}{2} R_{\mu\nu\rho\sigma} R^{\mu\nu\rho\sigma}
-1096\Lambda^2.
\end{eqnarray}
Even for a constant curvature background,
\begin{equation}
R_{\mu\nu\rho\sigma}=
(\Lambda/3)(g_{\mu\nu}g_{\rho\sigma}-g_{\mu\rho}g_{\nu\sigma}),\qquad
R_{\mu\nu\rho\sigma}R^{\mu\nu\rho\sigma}=\tfrac{8}{3}\Lambda^{2}\, ,
\end{equation}
there is no cancellation.  This demonstrates that, at the quantum level,
there remains a distinction between a pure massless spin-3/2 field and the
massless limit of a massive one.  The difference in the $b_4$
coefficients may be completely attributed to the presence of a single 
additional St\"uckelberg spinor generating the additional degrees of 
freedom for a massive field.  Note also that by subtracting the 
$b_{4}$ for opposite chiralities we recover the result that the 
gravitational axial anomaly for a spin $3/2$ particle is $-21$ times 
that of a spin $1/2$ particle in the massless case, but $-20$ times in 
the massless limit of the massive case \cite{CD1,CD2}, a result that is 
insensitive to the presence of the cosmological constant.

Note, however, that these discontinuities are essentially only one between
theories with different numbers of degrees of freedom.  In a
spontaneously broken supergravity theory, where the gravitino picks up a
mass, it does so by eating a Goldstino field that would be recovered in
the unbroken limit.  In this case no such discontinuity arises, since
the original theory necessarily contains spin-1/2 degrees of freedom in
addition to the gravitino, one combination of which will eventually
provide the appropriate Goldstino combination when supersymmetry is
broken.


\section{Massive supermultiplets}
\label{super}

The above computation of the $b_4$ coefficient for massive spin-3/2,
along with the spin-2 results of \cite{dilkes} now allow us to
complete the picture for all spins $\le2$.  For AdS representations
$D(E_0,s)$ with $s\le2$, the formal expressions for the one-loop
partition functions are given in Table~\ref{tbl1} for the massive case
and Table~\ref{tbl2} for the massless case.  We note that, for spins
$s=1$, $3/2$ and $2$, these expressions are compatible with the AdS
Higgs mechanism, (\ref{eq:adshig}), in the sense that no discontinuity
arises in the massless limit, provided the proper spin $s-1$ Goldstone
field is included.

Following the discussion of the previous section, we may determine 
the appropriate $b_4$ coefficients for both massive and massless 
fields in the Einstein background.  The resulting coefficients are 
given in Table~\ref{tbl3}.  This now allows us to examine some 
implications for massive supermultiplets.  We begin with $N=1$ AdS 
supergravity, with supergroup $OSp(1|4)$.  Excluding the 
supersingleton, we may evaluate the contribution to the $b_4$ 
coefficients from the various supermultiplets.  For the Wess-Zumino 
multiplet
\begin{equation}
{\cal D}(E_0,0)=D(E_0,0)+D(E_0+\half,\half)+D(E_0+1,0),
\end{equation}
we find
\begin{equation}
180(4\pi)^2b_4(E_0,0) =
\frac{15}{4}R_{\mu\nu\rho\sigma}R^{\mu\nu\rho\sigma}
+\Lambda^2[-20+60(E_0-1)^2],
\end{equation}
while for massless multiplets
\begin{equation}
{\cal D}(s+1,s)=D(s+1,s)+D(s+\ft32,s+\half)\qquad(s>0),
\end{equation}
we have

\begin{equation}
180(4\pi)^2b_4(s+1,s)=\cases{
-\frac{45}{4}R_{\mu\nu\rho\sigma}R^{\mu\nu\rho\sigma}-60\Lambda^2
\bstrut& $s=\fft12$ (Maxwell),\cr 
-\frac{285}{4}R_{\mu\nu\rho\sigma}R^{\mu\nu\rho\sigma}+500\Lambda^2
\bstrut&$s=1$ (gravitino),\cr 
\frac{615}{4}R_{\mu\nu\rho\sigma}R^{\mu\nu\rho\sigma}-1540\Lambda^2
\bstrut&$s=\fft32$ (graviton).\cr }
\end{equation}
More illuminating, perhaps, is the $E_0$ dependence of the $b_4$ coefficient
for a generic massive $N=1$ multiplet
\begin{eqnarray}
&{\cal D}(E_0,s)=D(E_0,s)+D(E_0+\half,s+\half)+D(E_0+\half,s-\half)
+D(E_0+1,s)&\nonumber\\
&(E_0>s+1; s>0).&
\end{eqnarray}
For spins less than two, we find
\begin{equation}
180(4\pi)^2b_4(E_0,s)=\cases{
-\frac{15}{2}R_{\mu\nu\rho\sigma}R^{\mu\nu\rho\sigma}
+\Lambda^2[-20-120(E_0-\half)(E_0-\ft32)]\kern-2cm\cr
\bstrut&$s=\fft12$ (massive vector),\cr
-\frac{315}{4}R_{\mu\nu\rho\sigma}R^{\mu\nu\rho\sigma}
+\Lambda^2[240+180E_0(E_0-2)]\kern-1cm\cr
\bstrut&$s=1$ (massive spin-$\fft32$),\cr
75R_{\mu\nu\rho\sigma}R^{\mu\nu\rho\sigma}
+\Lambda^2[-760-240(E_0+\half)(E_0-\ft52)]\kern-2cm\cr
\bstrut&$s=\fft32$ (massive spin-2).\cr
}
\end{equation}
In particular, potential terms proportional to $E_0^3$ and $E_0^4$ are
absent in the above.  This indicates that $N=1$ supersymmetry provides a
partial cancellation in the one-loop partition function, and furthermore
suggests that complete cancellation may be obtained in $N$-extended
supersymmetry.

This possibility for a cancellation among $b_4$ coefficients is most
readily seen in the case of $N=8$ AdS supergravity.  To see this, we
first recall that the one-loop divergences related to the $b_4$
coefficient may be canceled by the introduction of counterterms \cite{duff}
\begin{equation}
\gamma^{(s)}=\int d^4x \sqrt{g}\, b_4^{(s)}
=A \chi + B \delta,
\end{equation}
where $\chi$ is the Euler character and
$\delta=\frac{1}{12 \pi^2}\int d^4x\sqrt{g}$.  The same combinations
determine the trace anomaly. It was shown in \cite{Gibbons} that, for
the massless $SO(8)$ gauged $N=8$ supergravity multiplet, the total $B$
coefficient vanishes, {\it i.e.}~the cosmological constant $\Lambda$ is
not renormalized.  Since $\Lambda$ is related to the $SO(8)$ coupling
constant $e$ by $3e^{2}= -8\pi G\Lambda$, where $G$ is Newton's
constant, this implies a vanishing one loop beta function $\beta(e)$.

It was further shown in \cite{GN,Inami}, using zeta function methods,
that the vanishing of $\beta(e)$ continues to hold even for the massive
Kaluza-Klein tower arising from the round seven-sphere compactification
of eleven dimensional supergravity.  Having obtained in this paper the
massive spin $3/2$ $b_{4}$ coefficient in the presence of a cosmological
constant, we now have the complete set for spins $0\leq s \leq 2$, as
given in Table \ref{tbl3}.  So we can perform a novel calculation of the
beta function directly from the $b_{4}$ coefficients.  Using the well
known spectrum shown in Table~\ref{tbl4} \cite{englert,Casher:1984ym},
and the $b_4$ coefficients of Table~\ref{tbl3}, we obtain (at level $n$)
\begin{equation}
180(4\pi)^2b_4^{(N=8)}=-450\,d(n,0,0,0)R_{\mu\nu\rho\sigma}R^{\mu\nu\rho\sigma}
\label{eq:n8ans}
\end{equation}
(valid for $n\ge0$) where
\begin{eqnarray}
d(a_1,a_2,a_3,a_4)&=&(1+a_1)(1+a_2)(1+a_3)(1+a_4)
(1+\fft{a_1+a_2}2) (1+\fft{a_3+a_2}2) (1+\fft{a_4+a_2}2)\nonumber\\
&&(1+\fft{a_1+a_3+a_2}3) (1+\fft{a_1+a_4+a_2}3) (1+\fft{a_3+a_4+a_2}3)
\nonumber\\
&&(1+\fft{a_1+a_3+a_4+a_2}4) (1+\fft{a_1+a_3+a_4+2a_2}5)
\end{eqnarray}
is the Weyl formula for the dimension of the $SO(8)$ representation
given by Dynken label $(a_1,a_2,a_3,a_4)$.

This result confirms that the total beta function vanishes, level by
level, for the entire massive Kaluza-Klein tower.  In fact, recalling
that the massive tower is obtained by tensoring the $(n,0,0,0)$
representation (related to appropriate spherical harmonics on $S^7$)
with the massless supergraviton multiplet and Higgsing the result,
(\ref{eq:n8ans}) is naturally interpreted as a contribution from
$d(n,0,0,0)$ copies of the massless multiplet, each one of which carries
$180(4\pi)^2b_4=-450 R_{\mu\nu\rho\sigma}R^{\mu\nu\rho\sigma}$.


\section{Conclusion}
\label{conclusion}

We have seen that a cosmological constant cannot cure the VVZ problem
for spin $3/2$ at the quantum level since loop diagrams care that
a massive gravitino has four degrees of freedom and not two. This is in
keeping with a similar result for spin $2$ \cite{dilkes}.  

An interesting question is whether these spin $2$ and spin $3/2$ 
discontinuities appear in the Karch-Randall braneworld \cite{Karch}.  
According to \cite{Katz}, the spin $2$ (and, by implication, the spin 
$3/2$) discontinuity is absent.  If we try to analyze this from a 
brane, as opposed to bulk, perspective we see that the graviton 
acquires a mass by eating a massive spin one bound state of the CFT 
living on the brane \cite{Porrati:2001db}, a phenomenon peculiar to 
AdS.  In the supersymmetric case, the whole graviton supermultiplet 
acquires a mass by eating a supermultiplet of bound states of the SCFT 
($N=4$ Yang-Mills in the maximally supersymmetric case) \cite{Duff:2002ab}.  
However, the massless limit is much more subtle here because this is 
also the limit in which the AdS brane becomes Minkowski and for which, 
therefore, the bound states no longer appear.  Moreover, the 
propagators are no longer simply given by the inverses of the $\Delta$ 
operators given in \cite{dilkes} and in section \ref{massive} of this 
paper.  We intend to return to this issue elsewhere.

\section*{Acknowledgments}
This research was supported in part by DOE Grant DE-FG02-95ER40899.
We have greatly benefited from the help of F.A.~Dilkes, who collaborated
in much of the early work.
We wish to thank M.~Porrati for clarifying the connection between
the St\"uckelberg method and the super-Higgs mechanism.  JTL
acknowledges the hospitality of the KITP and the UCSB Physics Department
where part of this work was done.  MJD is grateful for correspondence
with S.M.~Christensen.


\begin{table}
\begin{tabular}{cl}
spin&massive partition function\\
\hline
0&$\det^{-\fft12}[\Delta(0,0)+E_0(E_0-3)]$\\
1/2&$\det^{\fft12}[\Delta_{(1/2)}+(E_0-\ft32)^2]$\\
1&$\det^{\fft12}[\Delta(0,0)+(E_0-1)(E_0-2)]
\det^{-\fft12}[\Delta\halfhalf+(E_0-1)(E_0-2)]$\\
3/2&$\det^{\fft12}[\Delta_{(3/2)}+(E_0-\ft32)^2]
\det^{-\fft12}[\Delta_{(1/2)}+(E_0-\ft32)^2+3]$\\
2&$\det^{\fft12}[\Delta\halfhalf+E_0(E_0-3)+6]
\det^{-\fft12}[\Delta(1,1)+E_0(E_0-3)+6]$
\end{tabular}
\caption{One loop partition functions for massive fields of spins
$\le2$.  The $E_0$ values are given in units of $\overline{m}$.}
\label{tbl1}
\end{table}

\begin{table}
\begin{tabular}{ccl}
spin&$E_0$&massless partition function\\
\hline
0&1 or 2&$\det^{-\fft12}[\Delta(0,0)-2\overline{m}^2]$\\
1/2&3/2&$\det^{\fft12}[\Delta_{(1/2)}]$\\
1&2&$\det[\Delta(0,0)]\det^{-\fft12}[\Delta\halfhalf]$\\
3/2&5/2&$\det^{\fft12}[\Delta_{(3/2)}+\overline{m}^2]
\det^{-1}[\Delta_{(1/2)}+4\overline{m}^2]$\\
2&3&$\det[\Delta\halfhalf+6\overline{m}^2]
\det^{-\fft12}[\Delta(1,1)+6\overline{m}^2]
\det^{-\fft12}[\Delta(0,0)+6\overline{m}^2]$
\end{tabular}
\caption{One loop partition functions for massless fields.  Gauge
invariance leads to a modified ghost structure for spins 1, 3/2 and 2
compared to Table~\ref{tbl1}.  The fermions are taken to be Dirac.}
\label{tbl2}
\end{table}

\begin{table}
\begin{tabular}{cll}
spin&&$180(4\pi)^2b_4$\\
\hline
0&&$R_{\mu\nu\rho\sigma}^2+\Lambda^2[36+40E_0(E_0-3)+10E_0^2(E_0-3)^2]$\\
1/2&&$-\ft74R_{\mu\nu\rho\sigma}^2+\Lambda^2[12-40(E_0-\ft32)^2
+20(E_0-\ft32)^4]$\\
1&$E_0=2$&$-13R_{\mu\nu\rho\sigma}^2-48\Lambda^2$\\
&$E_0>2$&$-12R_{\mu\nu\rho\sigma}^2+\Lambda^2[-132+30E_0^2(E_0-3)^2]$\\
3/2&$E_0=5/2$&$\ft{233}4R_{\mu\nu\rho\sigma}^2-548\Lambda^2$\\
&$E_0>5/2$&$\ft{113}2R_{\mu\nu\rho\sigma}^2+\Lambda^2[-96-320(E_0-\ft32)^2
+40(E_0-\ft32)^4]$\\
2&$E_0=3$&$212R_{\mu\nu\rho\sigma}^2-2088\Lambda^2$\\
&$E_0>3$&$200R_{\mu\nu\rho\sigma}^2+\Lambda^2[2460-1000(E_0^2-3E_0+6)
+50(E_0^2-3E_0+6)^2]$\\
\end{tabular}
\caption{$b_4$ coefficients for fields of spins $\le2$.  Note
that here the fermions are taken to be Majorana.}
\label{tbl3}
\end{table}

\begin{table}
\begin{tabular}{lll}
spin&$SO(8)$ irrep&$E_0$\\
\hline
$2^+$&$(n,0,0,0)$&$3+n/2$\\
$\fft32^{(1)}$&$(n,0,0,1)$&$\fft52+n/2$\\
$\fft32^{(2)}$&$(n-1,0,1,0)$&$\fft72+n/2$\\
$1^{-(1)}$&$(n,1,0,0)$&$2+n/2$\\
$1^+$&$(n-1,0,1,1)$&$3+n/2$\\
$1^{-(2)}$&$(n-2,1,0,0)$&$4+n/2$\\
$\fft12^{(1)}$&$(n+1,0,1,0)$&$\fft32+n/2$\\
$\fft12^{(2)}$&$(n-1,1,1,0)$&$\fft52+n/2$\\
$\fft12^{(3)}$&$(n-2,1,0,1)$&$\fft72+n/2$\\
$\fft12^{(4)}$&$(n-2,0,0,1)$&$\fft92+n/2$\\
$0^{+(1)}$&$(n+2,0,0,0)$&$1+n/2$\\
$0^{-(1)}$&$(n,0,2,0)$&$2+n/2$\\
$0^{+(2)}$&$(n-2,2,0,0)$&$3+n/2$\\
$0^{-(2)}$&$(n-2,0,0,2)$&$4+n/2$\\
$0^{+(3)}$&$(n-2,0,0,0)$&$5+n/2$
\end{tabular}
\caption{The spectrum of supergravity on the round seven-sphere.  For
$n=0$ or $1$, states with negative labels are not present.}
\label{tbl4}
\end{table}


\end{document}